\documentstyle[epsf]{mn2e}

\title [High Energy spectra of Seyferts and Unification schemes]
{High Energy X-ray spectra of Seyferts and Unification schemes for AGN}

\author[M.Middleton, C. Done, N. Schurch]
{Matthew Middleton$^1$, Chris Done$^1$ and Nick Schurch$^1$\\
$^1$Department of Physics, University of Durham, South Road, Durham
DH1 3LE,
UK\\
}

\date{}

\def\sax{{\it BeppoSAX\/}}
\def\inte{{\it INTEGRAL\/}}

\def\H0{{\rm ~km~s^{-1}~Mpc^{-1}}}

\def\etal{et al.~\/}

\def\la{\mathrel{\hbox{\rlap{\hbox{\lower4pt\hbox{$\sim$}}}{\raise2pt\hbox{$<$}}}}}
\def\ga{\mathrel{\hbox{\rlap{\hbox{\lower4pt\hbox{$\sim$}}}{\raise2pt\hbox{$>$}}}}}
\def\ls{\mathrel{\hbox{\rlap{\hbox{\lower4pt\hbox{$\sim$}}}\hbox{$<$}}}}
\def\gs{\mathrel{\hbox{\rlap{\hbox{\lower4pt\hbox{$\sim$}}}\hbox{$>$}}}}
\def\d25{D$_{\rm 25}$}

\def\.25{0.25 keV\thinspace}

\def\Mdot{\hbox{$\dot M$}}

\pagerange{\pageref{firstpage}--\pageref{lastpage}} \pubyear{2006}

\begin{document}

\topmargin = -0.5cm

\maketitle

\label{firstpage}
\begin{abstract}

The Unified Model of AGN predicts the sole difference between Seyfert
1 and Seyfert 2 nuclei is the viewing angle with respect to an
obscuring structure around the nucleus.  High energy photons above 20
keV are not affected by this absorption if the column is Compton thin,
so their 30--100~keV spectra should be the same. However, the observed
spectra at high energies appear to show a systematic difference, with
Seyfert 1's having $\Gamma \sim $2.1 whereas Seyfert 2's are harder
with $\Gamma \sim$ 1.9. We estimate the mass and accretion rate of
Seyferts detected in these high energy samples and show that they span
a wide range in $L/L_{Edd}$.  Both black hole binary systems and AGN
show a correlation between spectral softness and Eddington fraction,
so these samples are probably heterogeneous, spanning a range of
intrinsic spectral indices which are hidden in individual objects by
poor signal-to-noise. However, the mean Eddington fraction for the
Seyfert 1's is higher than for the Seyfert 2's, so the samples are
consistent with this being the origin of the softer spectra seen in 
Seyfert 1's. We stress that high energy spectra alone are not
necessarily a clean test of Unification schemes, but that the
intrinsic nuclear properties should also change with $L/L_{Edd}$.

\end{abstract}

\begin{keywords} accretion, accretion discs, black hole -- Galaxies: Seyfert
\vspace{-7mm}
\end{keywords}

\section{Introduction}

The simplest version of the Unification model of Antonucci \& Miller
(1985) is that the central engine (black hole, its accretion disc and
broad line region: BLR) are the same in all AGN, but that this is
embedded in an obscuring torus.  The nucleus is seen directly only for
inclination angles which do not intersect this material, giving the
classic Seyfert 1 AGN signature of a strong and variable UV/X-ray
continuum and broad emission lines. Conversely, where the
obscuration is in the line of sight, these features are hidden, and
the presence of an AGN can only be inferred from the high excitation
lines produced in the narrow line region on much larger scales
(Seyfert 2's). A key piece of evidence for this scenario is the
detection of polarized broad emission lines in classic Seyfert 2
galaxies (most notably in NGC 1068), showing that the BLR is present,
but can only be seen via scattered light (e.g. Antonucci \& Miller
1985; Tran 1995, Heisler et al. 1997). This scattering medium filling
the 'hole' in the torus is also detected in transmission in Seyfert
1's through the partially ionised absorption signatures seen in soft
X-rays (e.g. Blustin et al 2005).

\begin{figure*}
\begin{center}
\begin{tabular}{l}
\leavevmode \epsfxsize=12cm \epsfbox{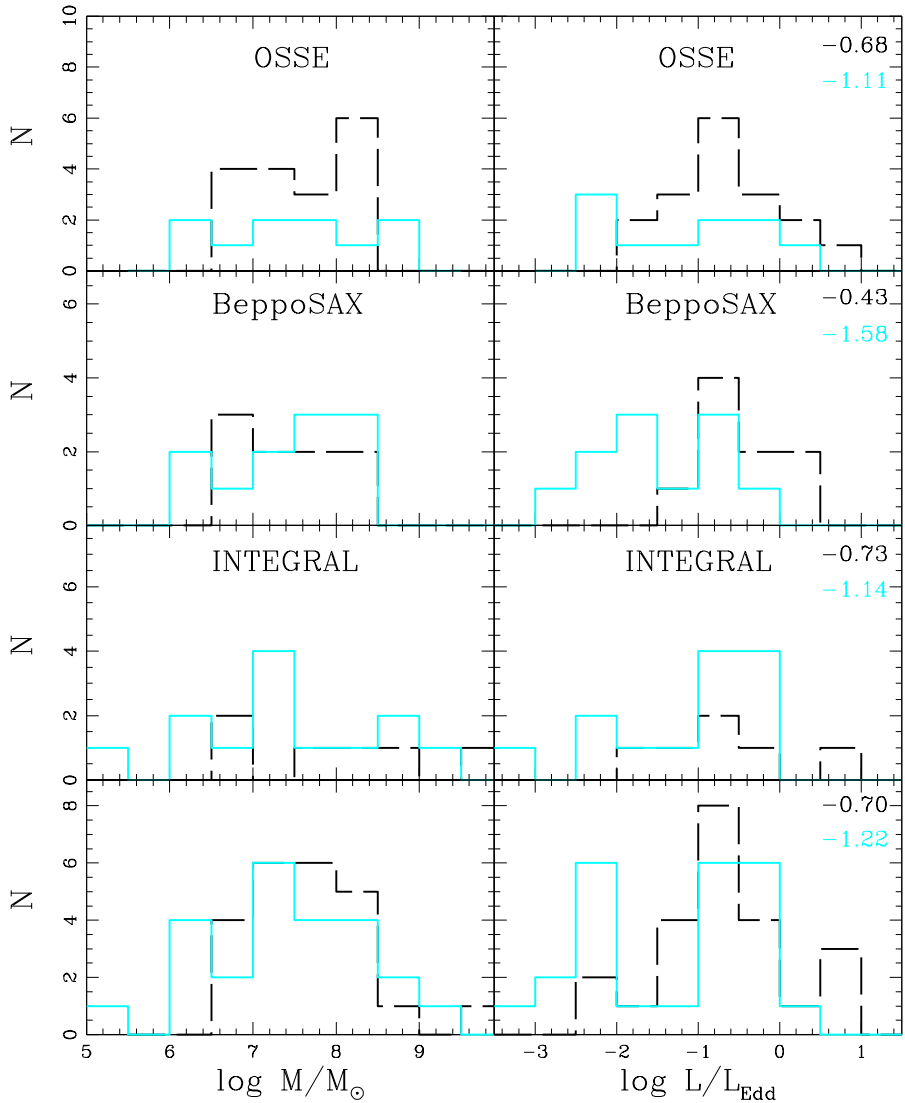}
\end{tabular}
\end{center}
\caption{Plots showing the distribution of log mass and log accretion rate for individual instruments and the full sample (bottom panels) with Seyfert 1 sources shown by a black dashed line and Seyfert 2 sources by a solid cyan line. The log mean accretion rate for each type is given in the top right hand corner.}
\label{}
\end{figure*}

The intrinsic X-ray spectrum of a type 1 AGN can be well described by
a power law (of photon spectrum $N(E)\propto E^{-\Gamma}$) together
with its Compton reflection from the optically thick accretion disc
and/or torus (e.g. Nandra \& Pounds 1994). Again, the Unified models
are supported by X-ray observations, which show that this intrinsic
spectrum is substantially suppressed at low energies due to absorption
in Seyfert 2's (Awaki et al 1991; Smith \& Done 1996; Turner et al
1997; Bassani et al 1999; Cappi et al 2006). However, the strong
energy dependence of photoelectric absorption means that this is
unlikely to affect samples above 10 keV for columns which are not
Compton thick, yet high energy experiments show that the intrinsic
spectra of Seyfert 2 galaxies are systematically harder ($\Gamma$
$\sim$ 1.9 - 2) than than the Seyfert 1's, which have $\Gamma > $ 2
(Zdiarski et al.  1995, Gondek et al. 1996, Perola et al. 2002, Malizia et al. 2003, Beckmann et al. 2006). This is not consistent with
the idea that these nuclei are identical, unless the intrinsic
emission is anisotropic, being harder in the equatorial plane. 

However, all accreting black hole systems (both stellar and
supermassive) generally show softer intrinsic spectra at higher
accretion rates (hereafter parameterised as Eddington fraction,
$L/L_{Edd}$), e.g. Laor (2000); Remillard \& McClintock (2006). Thus
inclination alone is not the sole determinant of the observed
spectrum, and Seyferts at different $L/L_{Edd}$ should not be expected
to have the same intrinsic emission. Here we collate estimates of
black hole mass and mass accretion rate for the Seyfert 1 and 2's
detected out to $>50$~keV from CGRO (OSSE), BeppoSAX (PDS) and
INTEGRAL (IBIS) instruments.  We show that these high energy samples
of Seyferts span a large range $L/L_{Edd}$, so should include a range
of intrinisic spectral slopes. The mean $L/L_{Edd}$ for the Seyfert
1's in the sample is higher than that for Seyfert 2's, consistent with
the steeper spectral slope inferred for the Seyfert 1's being due to
the intrinsic spectrum softening at higher $L/L_{Edd}$. We stress that
samples of Seyfert 1 and 2 AGN need to be matched on intrinsic
properties such as $L/L_{Edd}$ in order to explore differences in
orientation.

\begin{table*}
\begin{center}
\begin{minipage}{125mm} 
\bigskip
\begin{tabular}{ll|l|c|c}
  \hline
  Source & Instrument & logM/M$_{\odot}$ &  L/L$_{Edd}$   &Reference\\
 \hline

\vspace{1mm}MCG--6-30-15  & {\em OSSE }, {\em BeppoSAX}, {\em INTEGRAL} &  6.81 $^{+0.54}_{-0.54}$  $^{\sigma}$  &   0.25 $^{x}$    &  6 (18) + 23\\
\vspace{1mm}IC 4329A    & {\em OSSE }&   6.70 $^{+0.56}_{-peg}$ (mean) $^{r}$   & 0.79 $^{f}$  & 20+ 20\\
\vspace{1mm}&&6.85  $^{+0.55}_{-peg}$ (rms) $^{r}$ &&\\
\vspace{1mm}MR 2251-178& {\em INTEGRAL}&   6.92  $^{+0.22}_{-0.22}$ $^{m}$  &  4.68 $^{z}$          & 10 + 21 \\
\vspace{1mm}NGC3783    &   {\em OSSE }, {\em BeppoSAX}&  6.97  $^{+0.30}_{-0.97}$ (mean) $^{r}$  & 0.23 $^{f}$   &  2 + 20 \\
\vspace{1mm}&& 7.04  $^{+0.30}_{-0.96}$ (rms) $^{r}$ &&\\
\vspace{1mm}NGC7469  & {\em OSSE }, {\em BeppoSAX}  &    6.81  $^{+0.30}_{-peg}$ (mean) $^{r}$   &  2.14 $^{f}$    & 20 + 20\\
\vspace{1mm}&& 6.88  $^{+0.29}_{-peg}$ (rms) $^{r}$ &&\\
\vspace{1mm}ESO 141-55  & {\em OSSE }, {\em BeppoSAX}  &  7.10 $^{+0.57}_{-0.57}$ $^{m}$  &   2.51 $^{ox}$     & 1\\
\vspace{1mm}MCG--2-58-22 & {\em OSSE } &    7.14 $^{+0.57}_{-0.57}$  $^{m}$  &   3.47 $^{ox}$     & 1\\
\vspace{1mm}NGC6814     & {\em OSSE }   &    7.08  $^{+0.57}_{-0.57}$ $^{r}$   &   0.05 $^{f}$      & 22 + 2\\
\vspace{1mm}3C120    & {\em BeppoSAX}   &  7.36 $^{+0.22}_{-0.28} $ (mean)$^{r}$  & 0.65 $^{f}$         & 20 + 20 \\ 
\vspace{1mm}&&7.48 $^{+0.21}_{-0.27}$ (rms) $^{r}$  &&\\
\vspace{1mm}NGC3516    & {\em OSSE }    &    7.36   $^{+0.59}_{-0.59}$  $^{r}$  &  0.07 $^{f}$ & 2\\
\vspace{1mm}Mrk279     & {\em OSSE }   &    7.54 $^{+0.10}_{-0.13}$  $^{r}$  &  0.13 $^{z}$     & 5\\
\vspace{1mm}NGC3227    & {\em OSSE }    &    7.59 $^{+0.19}_{-peg}$ (mean) $^{r}$  &  0.01 $^{f}$      & 20 + 2\\
\vspace{1mm}&&7.69 $^{+0.18}_{-peg}$ (rms) $^{r}$  &&\\
\vspace{1mm}NGC5548 & {\em OSSE }, {\em BeppoSAX}    &   8.09   $^{+0.09}_{-0.07}$ (mean)  $^{r}$  &  0.05 $^{f}$  & 20 + 20\\
\vspace{1mm}&& 7.97 $^{+0.08}_{-0.07}$ (rms) $^{r}$  &&\\
\vspace{1mm}1H 1934-063 & {\em INTEGRAL}  &  7.86  $^{+0.63}_{-0.63}$  $^{m}$ &  0.07 $^{ox}$    &  1\\
\vspace{1mm}Fairall 9   & {\em BeppoSAX}   &  7.90 $^{+0.11}_{-0.31}$ (mean) $^{r}$   & 0.16 $^{s}$   &  20 + 20\\
\vspace{1mm}&&7.91 $^{+0.11}_{-0.32}$ (rms) $^{r}$  &&\\
\vspace{1mm}NGC7213    & {\em OSSE }   &    7.99  $^{+0.64}_{-0.64}$ $^{\sigma}$   &   0.02 $^{f}$        & 2\\
\vspace{1mm}MCG +8-11-11  & {\em OSSE } &    8.06  $^{+0.64}_{-0.64}$  $^{m}$   & 0.35 $^{ox}$           & 1\\
\vspace{1mm}NGC526A    & {\em OSSE }, {\em BeppoSAX}    &    8.11  $^{+0.65}_{-0.65}$ $^{\sigma}$   &  0.17 $^{ox}$     &  6 (18) + 1\\
\vspace{1mm}Mrk841     & {\em OSSE }   &    8.49  $^{+0.68}_{-0.68}$ $^{r}$  &  0.17 $^{f}$           & 22 + 2\\
\vspace{1mm}Mrk509    & {\em OSSE }, {\em BeppoSAX}, {\em INTEGRAL}    &    8.16 $^{+0.04}_{-0.04}$  $^{r}$  &  0.45 $^{ox}$    &  5 + 1\\
\vspace{1mm}III Zw 2    & {\em OSSE }    &  8.42  $^{+0.67}_{-0.67}$ $^{r}$ &  0.13 $^{z}$   & 3\\
\vspace{1mm}PG 1416-129  & {\em INTEGRAL}  &    8.75   $^{+0.70}_{-0.70}$ $^{r}$  & 0.12 $^{z}$     & 3\\
\vspace{1mm}3C111       & {\em INTEGRAL}  &  9.56  $^{+0.76}_{-0.76}$  $^{m}$&  0.01 $^{z}$    & 9\\

\hline

\end{tabular}

\end{minipage}

\end{center}
\caption{Seyfert 1 sub-sample. Name superscripts indicate the hard
X-ray satellite that observed the source, o: {\it OSSE}, i: \inte ~\&
b: \sax. Mass superscripts refer to the measurement method, $\sigma$:
stellar velocity dispersion, r: reverberation mapping, m:
other. Accretion rate superscripts denote the method by which the
bolometric luminosity was determined, x: correction from X-ray, f:
integration of the SED flux, ox: correction from O{\small III}, s: SED
modelling, z: other. References, 1: Wang \etal (2007), 2: Woo \& Urry
(2002), 3: Hao \etal (2005), 4: McHardy (1988), 5: Vestergaard \&
Peterson. (2006), 6: Garcia-Rissmann \etal (2005), 7: Bian \& Gu
(2007), 8: Greenhill \etal (1997), 9: Grandi \etal (2006), 10: Morales
\& Fabian (2002), 11: Marconi \etal (2006), 12: Whysong \& Antonucci
(2004),13: Awaki \etal (2005), 14: Czerny \etal (2001), 15: Tadhunter
\etal (2003), 16: van Bemmel \etal (2003), 17: Gu \etal (2006), 18:
Tremaine \etal (2002), 19: Risaliti \etal (2005), 20: Kaspi \etal
(2000), 21: Monier \etal (2001), 22: Laor \etal (2001) 23: Dadina
(2007). Where more than one reference is given, the first refers to
the black hole mass and the second to the accretion rate. Where
reference 18 is given in brackets, the relation from Tremaine \etal
(2002) has been used to determine the black hole mass via stellar
velocity dispersion.}

\end{table*}

\begin{table*}
\begin{center}
\begin{minipage}{110mm} 
\bigskip
\begin{tabular}{ll|l|c|c}
  \hline
  Source & Instrument & logM/M$_{\odot}$ &  L/L$_{Edd}$  & Reference\\
 \hline

\vspace{1mm}NGC6300    & {\em INTEGRAL}    &    5.45  $^{+0.44}_{-0.44}$ $^{m}$   &   0.91 $^{x}$    & 13\\
\vspace{1mm}NGC7314    & {\em BeppoSAX}   &    6.03  $^{+0.48}_{-0.48}$ $^{\sigma}$   &   0.41 $^{x}$    & 17 (+18)\\
\vspace{1mm}NGC4945 &  {\em OSSE }, {\em INTEGRAL}   &    6.15   $^{+0.49}_{-0.49}$ $^{\sigma}$  &  0.17 $^{x}$    & 8\\
\vspace{1mm}MCG--5-23-16  &  {\em OSSE }, {\em BeppoSAX} &    6.29 $^{+0.50}_{-0.50}$  $^{m}$  &   0.09 $^{ox}$     &  1\\
\vspace{1mm}Circinus     & {\em INTEGRAL}  &  6.42 $^{+0.51}_{-0.51}$  $^{\sigma}$ &  0.32 $^{ox}$      & 7\\
\vspace{1mm}NGC5506    &  {\em OSSE }   &    6.65  $^{+0.53}_{-0.53}$ $^{\sigma}$   &   2.51 $^{ox}$     & 7\\
\vspace{1mm}NGC4593   & {\em INTEGRAL}, {\em BeppoSAX}    &    6.91  $^{+0.55}_{-0.55}$ $^{r}$ &  0.12 $^{f}$    &2\\
\vspace{1mm}ESO 103-G35  & {\em INTEGRAL}, {\em BeppoSAX}  &  7.14 $^{+0.57}_{-0.57}$  $^{m}$ &  0.01 $^{x}$     & 14\\
\vspace{1mm}NGC4388    &  {\em OSSE }, {\em INTEGRAL}    &    7.22   $^{+0.58}_{-0.58}$ $^{\sigma}$  & 0.79 $^{ox}$     & 7\\
\vspace{1mm}NGC1068   & {\em INTEGRAL}   &    7.23  $^{+0.58}_{-0.58}$ $^{\sigma}$  &  0.44 $^{f}$       & 8 + 2\\
\vspace{1mm}NGC7582   &  {\em OSSE }    &    7.25   $^{+0.58}_{-0.58}$ $^{\sigma}$  & 0.35 $^{ox}$      & 7\\ 
\vspace{1mm}NGC5674    & {\em BeppoSAX}   &    7.36  $^{+0.59}_{-0.59}$ $^{\sigma}$   &  0.17 $^{x}$    &  17 (+18)\\
\vspace{1mm}ESO 323-G077  & {\em INTEGRAL} &  7.39 $^{+0.59}_{-0.59}$ $^{m}$  &  0.28 $^{ox}$    & 1\\
\vspace{1mm}NGC4507   &  {\em OSSE }, {\em INTEGRAL}    &    7.58   $^{+0.61}_{-0.61}$ $^{\sigma}$  &  0.27 $^{ox}$   & 7\\
\vspace{1mm}NGC4258    & {\em BeppoSAX}   &    7.62   $^{+0.61}_{-0.61}$ $^{\sigma}$  & 0.01 $^{f}$       & 2\\
\vspace{1mm}NGC7172    &  {\em OSSE }, {\em BeppoSAX}   &    7.67  $^{+0.61}_{-0.61}$ $^{\sigma}$   &   3x10$^{-3}$ $^{ox}$   & 7\\
\vspace{1mm}NGC2992   & {\em BeppoSAX}   &    7.72  $^{+0.62}_{-0.62}$ $^{\sigma}$  &  0.01 $^{f}$   &  2\\  
\vspace{1mm}NGC5252    & {\em BeppoSAX}   &    8.04  $^{+0.64}_{-0.64}$  $^{\sigma}$  &  0.17 $^{s}$   & 2\\
\vspace{1mm}Centaurus A  & {\em INTEGRAL}  &  8.04 $^{+0.64}_{-0.64}$ $^{\sigma}$  &  7x10$^{-4}$ $^{s}$   &  11 + 12\\
\vspace{1mm}NGC1365    & {\em BeppoSAX}  &    8.18  $^{+0.65}_{-0.65}$ $^{m}$  &  2x10$^{-3}$ $^{x}$      & 19\\
\vspace{1mm}NGC2110    &  {\em OSSE }, {\em BeppoSAX}    &    8.30  $^{+0.66}_{-0.66}$ $^{\sigma}$  &  5x10$^{-3}$ $^{f}$    &  2\\
\vspace{1mm}NGC1275    &  {\em OSSE }, {\em INTEGRAL}    &    8.51  $^{+0.68}_{-0.68}$ $^{\sigma}$  &  0.03 $^{f}$   &  2\\
\vspace{1mm}Mrk3       &  {\em OSSE }, {\em INTEGRAL}    &    8.65 $^{+0.69}_{-0.69}$  $^{\sigma}$  &  6x10$^{-3}$ $^{f}$    &  2\\
\vspace{1mm}Cyg A      & {\em INTEGRAL}   &  9.40 $^{+0.75}_{-0.75}$  $^{\sigma}$ &  8x10$^{-3}$ $^{x}$    & 15 + 16\\

\hline

\end{tabular}

\end{minipage}

\end{center}
\caption{Seyfert 2 sub-sample.  Superscripts and references as for Table 1.}

\end{table*}

\section{Data}

We construct a moderately sized sample of local hard X-ray detected
AGN. These data, shown in Table 1 \& 2, are taken from Malizia \etal
(2003 - \sax), Zdziarski \etal (2000 - {\it
OSSE}) and Beckmann \etal (2006 - \inte), resulting in a total sample
of 47 AGN; 23 Seyfert 1s and 24 Seyfert 2s. There is always some
ambiguity in assigning objects as type 1 or 2, firstly as there are
optical intermediate types, and secondly as the absorption environment
is complex. There is (Compton thin) molecular gas in the plane of the
galaxy as well as the much smaller scale (and possibly Compton thick)
molecular torus (Maiolino \& Rieke 1995), and these two obscuring
structures are not aligned (Nagar \& Wilson 1999; Schmitt et al
2002). Thus a face on (type 1) nucleus can be obscured at both optical
and X--ray wavelengths, leading to a type 2 classification
(e.g. NGC~5506: Nagar et al 2002).  Nonetheless, these classifications
are those for which the difference in X-ray spectral index is claimed,
so this is the only sample definition we can use to explore whether
accretion rate can be the reason for this difference.

The black hole masses for the objects in the sample are derived
primarily from stellar velocity dispersion (21 objects) and
reverberation mapping (15 objects).  For the remaining 11 sources the
black hole masses are inferred from optical line widths and X-ray
variability measurements. For those sources with reverberation mapping
measurements, the black hole mass errors quoted are those stated in
the referenced literature. For the remaining sources, the quoted black
hole mass errors are calculated from the relation presented in
Tremaine \etal (2002).

The bolometric luminosities for the majority of the sample (22
objects) have been calculated by converting the O{\small III} and/or
X-ray luminosity to bolometric luminosity, using the relations given
in Wang \& Zhang (2007) and Satyapal et al. (2005) respectively. For
the remaining sources, 20 have bolometric luminosities measured from
the AGN SED (17 from integration of the SED \& 3 from modelling) and
the remaining 5 sources have bolometric luminosities stated in the
literature. The mass accretion rate is parameterized by the ratio of
the bolometric luminosity to the Eddington luminosity for the given
black hole mass, i.e. \Mdot=$L_{bol}/L_{Edd}$ where $L_{Edd} =
1.3\times 10^{38} (M_{BH}/M_{\odot})$.

Fig 1 shows the distribution of black hole mass (left panel) and
Eddington fraction (right panel) for each of the high energy samples
of Seyfert 1's (black, dashed line) and Seyfert 2's (cyan, solid
line). It is plain that the mass accretion rates span a wide range
in  $L/L_{Edd} \sim 10^{-3}$--1. This means that these AGN form a
very heterogeneous sample, with the clear expectation that their high
energy spectra are also heterogeneous. 

The mean log mass accretion rate for each of the samples is given in
the top right hand corner of the plot with Seyfert 1 in black and
Seyfert 2 in red. All three instruments have higher mean $L/L_{Edd}$
for Seyfert 1's than for the Seyfert 2's. For the complete sample we
perform a series of 10,000 Monte Carlo Bootstrap simulations in order
to determine rigorously the statistical significance of differences
between the Seyfert 1 and Seyfert 2 distributions. The Monte Carlo
Bootstrap simulations randomly select objects from the parent sample
to construct paired random sub-samples (of the correct size). We then
calculate the mean of the interesting parameters for each randomly
generated sub-sample, and the difference between these means for the
sub-samples in each pair. We then measure the frequency with which our
random sub-samples show differences greater than, or equal to the
differences measured between the Seyfert 1 and Seyfert 2 sub-samples.

The result of the bootstrap analysis shows that the difference is
significant at only $\sim$1.7$\sigma$ for the combined sample, though
the mean $L/L_{Edd}$ for the Seyfert 2's and Seyfert 1's are $0.06$
and $0.20$, respectively. We also try to assess the systematic
uncertainties in this result due to the mass determination.  H$\beta$
FWHM values can give an alternative measure of black hole mass for the
Seyfert 1's (e.g. Salviander \etal 2006), and we find that these
are within 0.3 dex for all but one source (MR 2251-178). Removing all
sources where the mass and resulting accretion rate is not found from
stellar velocity dispersion or reverberation mapping (which removes MR
2251-178) reduces the significance of the difference to $<$1$\sigma$ due to the smaller sample size. Nevertheless, the Seyfert 1 sample still has the
higher mean accretion rate (0.15 compared to 0.06 for the Seyfert
2's).

\section{Discussion}

The X-ray spectra of stellar mass black hole binaries can be well
described by a combination of emission from the accretion disc and a
Comptonised tail (see e.g. the reviews by Remillard \& McClintock
2006; Done, Gierlinski \& Kubota 2007). The relative importance of the
disc and tail can change dramatically, as does the shape of the tail,
giving rise to the well known 'spectral states' as a function of
(average) $L/L_{Edd}$. In the low/hard state, generally seen at
luminosities below a few percent of Eddington, the tail dominates the
X-ray emission and its spectral index is fairly well correlated with
(average) mass accretion rate, with $\Gamma\sim 1.5$ at the lowest
luminosities softening to $\Gamma\sim 2.1$ just before the major
transition to the soft, or disc dominated states. The tail in these
states can be very weak, carrying less than 20 per cent of the total
bolometric power, when it has $\Gamma\sim 2.1$ (disc dominant
state). Conversely, the tail can also be much stronger and softer,
with $\Gamma\sim 2.5$ (very high state or steep power law state).

AGN spectra should show analogous spectral states if the properties of
the accretion flow scale simply with black hole mass. There is
evidence for this simple scaling, from the radio-X-ray `fundamental
plane' relations (Merloni, Heinz, \& Di Matteo 2003; Falcke, K\"ording
\& Markoff 2004) and X-ray variability properties (McHardy et al. 2007;
Gierlinski et al. 2007). In this picture (see e.g. Boroson 2002),
LINERS, with their hard X-ray spectra and weak UV disc, are the
analogues of the low/hard state (but see Maoz 2007), while Seyfert 1
and QSO's at higher $L/L_{Edd}$ correspond to the disc dominated
states, with the characteristic strong UV disc emission and consequent
broad emission lines, and the Narrow line Seyfert 1's at the highest
$L/L_{Edd}$ may be the analogues of the very high state (Pounds, Done
\& Osborne 1995; Middleton, Done \& Gierli\'nski 2007).

Observations of type 1 AGN spectra in the 2--10~keV band are
consistent with this predicted softening of the intrinsic spectrum
with increasing $L/L_{Edd}$ (e.g. Laor 2000), so it seems highly
likely that the high energy X--ray spectra considered here should also
change with $L/L_{Edd}$. We show that current samples of Seyferts
detected at high energies are clearly highly heterogeneous in this
parameter (see Fig 1), so the mean spectral index will be determined
by a signal--to--noise weighted average of $L/L_{Edd}$. The
$L/L_{Edd}$ distributions for the total Seyfert 1 and Seyfert 2
samples presented here are not significantly different (statistically)
however the individual OSSE, BeppoSAX and INTEGRAL samples considered
here are all consistent with having a higher mean $L/L_{Edd}$ for the
Seyfert 1's than for the Seyfert 2s. Thus, despite the limited sample
size the apparently steeper spectra seen in the Seyfert 1's at high
energy may be be due to an intrinsic softening with increasing
$L/L_{Edd}$. Future studies with larger samples of high energy spectra
of AGN should be able to test unambiguously whether $L/L_{Edd}$ is the
major parameter in determining the shape of the 20--100~keV spectrum
in Compton thin Seyferts.

\label{lastpage}

\end{document}